\documentclass[superscriptaddress,nofootinbib,a4paper,11pt]{article}
\usepackage{jheppub}
\usepackage{graphicx}
\usepackage{subcaption}
\usepackage{float}
\usepackage{amsmath}
\usepackage{amssymb}
\usepackage[utf8]{inputenc}
\usepackage{hyperref}
\usepackage{color}
\usepackage{slashed}
\usepackage{multirow}
\usepackage{enumitem}
\usepackage{lineno}
\allowdisplaybreaks

\begin{document}

\title{\boldmath Excluding Local Hidden Variables in  $\Lambda\bar{\Lambda}$ Production: The Incompatibility with Angular-Momentum Conservation and CPT Invariance}

\author[a,*]{Junle Pei,\note[*]{Corresponding authors.}}
\author[b,*]{Lina Wu,}
\author[c,*]{Tianjun Li,}
\author[c]{Xiqing Hao}

\affiliation[a]{Institute of Physics, Henan Academy of Sciences, Zhengzhou 450046, P. R. China}
\affiliation[b]{School of Sciences, Xi'an Technological University, Xi'an 710021, P. R. China}
\affiliation[c]{School of Physics, Henan Normal University, Xinxiang 453007, P. R. China}

\emailAdd{peijunle@hnas.ac.cn}
\emailAdd{wulina@xatu.edu.cn}
\emailAdd{tli@itp.ac.cn}
\emailAdd{haoxiqing@htu.edu.cn}

\abstract{
We analyze spin entanglement in $\Lambda\bar{\Lambda}$ pairs produced in the decays of spin-zero particles, contrasting predictions from quantum field theory (QFT) with those of local hidden-variable theories (LHVTs). Using the self-analyzing weak decays $\Lambda \to p\pi^-$ and $\bar{\Lambda} \to \bar{p}\pi^+$, we derive the joint angular distributions within QFT. Our key findings are: For scalar production $h \to \Lambda\bar{\Lambda}$, no LHVT respecting locality and angular-momentum conservation can reproduce the QFT distribution. For pseudoscalar production $a \to \Lambda\bar{\Lambda}$, a CPT-symmetric LHVT is excluded by positivity constraints given the measured analyzing powers; however, if CPT symmetry is relaxed, an explicit LHVT construction---with uniform hidden-variable measure and response functions satisfying $b_1 c_1 = 3\alpha_{\Lambda}\alpha_{\bar{\Lambda}}$---can match the QFT result. For the most general spin-zero decay $s\to \Lambda\bar{\Lambda}$ with arbitrary scalar-pseudoscalar mixing, we, under CPT invariance, identify the regions of parameter space where the QFT joint angular distribution does or does not admit an LHVT realization. These distinct signatures provide clear, experimentally testable criteria to discriminate between QFT and LHVT in $\Lambda\bar{\Lambda}$ systems across different production mechanisms.
}

\maketitle

\section{Introduction} \label{intro} 

The conflict between quantum mechanics and local hidden-variable theories (LHVTs) has been a central problem in foundational physics since the seminal work of Einstein, Podolsky, Rosen~\cite{PhysRev.47.777}, and Bell~\cite{PhysicsPhysiqueFizika.1.195}. While extensive experimental tests have been conducted with photons~\cite{PhysRevLett.28.938,PhysRevLett.47.460,PhysRevLett.49.1804,Bouwmeester_1997,PhysRevLett.81.3563,PhysRevLett.80.1121}, atoms~\cite{Riebe:2004jpa,Barrett:2004bxh}, leptons~\cite{Hensen:2015ccp,Han:2025ewp}, quarks~\cite{Afik:2020onf,Fabbrichesi:2021npl,Han:2023fci,ATLAS2024,Cheng:2023qmz,CMS:2024pts,Cheng:2025cuv}, and gauge bosons~\cite{Subba:2024aut}, the baryon sector---particularly hyperon-antihyperon pairs~\cite{BESIII:2018cnd,Pei:2025yvr,BESIII:2025vsr,Lin:2025eci,Wu:2025dds} such as $\Lambda\bar{\Lambda}$---remains a critical yet underexplored domain for probing quantum nonlocality and fundamental symmetries.  

In this study, we first focus on the spin entanglement of $\Lambda\bar{\Lambda}$ pairs produced in the decays of scalar and pseudoscalar particles. These two production mechanisms lead to distinct spin correlations that are deeply rooted in the transformation properties of the parent particle under Lorentz and discrete symmetries. The joint angular distributions of the decay protons and antiprotons serve as the primary observables, encoding the spin structure of the $\Lambda\bar{\Lambda}$ system and enabling a direct comparison between the predictions of quantum field theory (QFT) and those of LHVTs.

Inspired by the discussions in Refs.~\cite{ABEL1992304,Li:2024luk,Bechtle:2025ugc,Abel:2025skj,Low:2025aqq}, a central aim of this work is to delineate the extent to which LHVTs can mimic the spin correlations predicted by QFT in these processes. We systematically explore the constraints imposed by locality, angular-momentum conservation, and CPT symmetry on possible LHVT realizations. Our analysis reveals a striking difference between the scalar and pseudoscalar cases: while the scalar channel inherently resists any local hidden-variable description, the pseudoscalar channel admits such a description only at the cost of abandoning CPT invariance. This contrast not only sharpens the distinction between quantum and classical descriptions of spin entanglement but also opens new avenues for probing fundamental symmetries in baryonic systems.

We further extend the analysis to the most general spin-zero decay $s\to \Lambda\bar{\Lambda}$, allowing arbitrary complex scalar and pseudoscalar couplings. This general case includes the pure scalar and pure pseudoscalar scenarios as special limits, but it also covers spin-zero parents with no definite parity assignment. With CPT invariance imposed, we map out the parameter space in which an LHVT realization of the QFT joint angular distribution is allowed or excluded.

The structure of this paper is as follows. In Sec.~\ref{sec:2}, we establish the theoretical framework for the QFT predictions, deriving the joint angular distributions for the decays $h\to\Lambda\bar{\Lambda}$ and $a\to\Lambda\bar{\Lambda}$. Sec.~\ref{sec:3} is devoted to the construction of LHVTs compatible with angular-momentum conservation and locality. We prove the impossibility of reproducing the scalar decay distribution under any LHVT, and analyze the pseudoscalar case in detail---both with and without CPT symmetry---presenting an explicit model that matches the QFT result when CPT is violated. In Sec.~\ref{sec:spinzero}, we extend the analysis to the most general spin-zero decay $s\to\Lambda\bar{\Lambda}$ with arbitrary scalar-pseudoscalar mixing and, under the assumption of CPT invariance, delineate the regions of parameter space in which the QFT-predicted joint angular distribution does or does not admit an LHVT realization.
Finally, we conclude our study in Sec.~\ref{con}.

\section{Theoretical Framework for QFT Predictions} \label{sec:2} 

In line with Refs.~\cite{Pei:2025non,Pei:2025yvr,Pei:2025ito}, we use the process $\Lambda(\to p\pi^-)\bar{\Lambda}(\to \bar{p}\pi^+)$ as a concrete example.
For a $\Lambda\bar{\Lambda}$ system in a pure state, the most general polarization configuration in the helicity basis can be expressed as
\begin{align}
    \left| \Lambda \bar{\Lambda}\right\rangle=\sum_{k,j=\pm\frac{1}{2}} \alpha_{k,j} \left|k\right\rangle_\Lambda \left|j\right\rangle_{\bar{\Lambda}}~,
\end{align}
where $k$ and $j$ are the helicity quantum numbers of $\Lambda$ and $\bar{\Lambda}$, respectively, defined along their own momentum directions in the center-of-mass (c.m.) frame.
To set conventions, in the decays $\Lambda \to p+\pi^-$ and $\bar{\Lambda} \to \bar{p}+\pi^+$ we describe the unit three-momenta of $p$ (in the $\Lambda$ rest frame) and $\bar{p}$ (in the $\bar{\Lambda}$ rest frame) by spherical angles:
\begin{align}
& \hat{e}_{p/\bar{p}}=(\sin\theta_{1/2}\cos\phi_{1/2},\sin\theta_{1/2}\sin\phi_{1/2},\cos\theta_{1/2})~.    
\end{align}
The polar angles $\theta_i~(i=1,2)$ are measured with respect to the $\Lambda$ flight direction $\hat{e}_\Lambda$ in the $\Lambda\bar{\Lambda}$ c.m. frame. The azimuthal angles $\phi_i$ ($\phi_i\in [0,~2\pi]$ for $i=1,2$) are defined relative to an arbitrary axis orthogonal to $\hat{e}_\Lambda$ and increase according to the right-hand rule about $\hat{e}_\Lambda$.

Our strategy is to tomographically reconstruct the spin polarization of the $\Lambda\bar{\Lambda}$ pair from angular observables. Consequently, the fundamental input is the fully differential angular distribution of the final state, denoted $\mathcal{W}(\theta_1,\theta_2,\phi_1,\phi_2)$.
An explicit expression for $\mathcal{W}(\theta_1,\theta_2,\phi_1,\phi_2)$ is ~\cite{Pei:2025yvr}
\begin{align}
   \mathcal{W}(\theta_1,\theta_2,\phi_1,\phi_2)=&\frac{1}{16\pi^2}\sum_{\lambda_p,\lambda_{\bar{p}}=\pm \frac{1}{2}}\left(1-2\lambda_p\alpha_\Lambda\right)\left(1-2\lambda_{\bar{p}}\alpha_{\bar{\Lambda}}\right) \sum_{k,j,m,n=\pm \frac{1}{2}} \alpha_{k,j}\alpha_{m,n}^*\times
    \nonumber\\
  & e^{i(k-m)\phi_1} e^{i(n-j)\phi_2}
   d^{\frac{1}{2}}_{k,\lambda_p}(\theta_1)
   d^{\frac{1}{2}}_{m,\lambda_p}(\theta_1)
   d^{\frac{1}{2}}_{\lambda_{\bar{p}},j}(\pi-\theta_2)
   d^{\frac{1}{2}}_{\lambda_{\bar{p}},n}(\pi-\theta_2)~.
\end{align}
Here, \(\lambda_{p}\) and \(\lambda_{\bar{p}}\) ($\lambda_p,\lambda_{\bar{p}}=\pm\frac{1}{2}$) are spin projections of $p$ and $\bar{p}$ defined relative to directions of \(\hat{e}_p\) and \(\hat{e}_{\bar{p}}\), respectively, $\alpha_{\Lambda/\bar{\Lambda}}$ are spin analzing powers, and $d_{k,\lambda_p}^{\frac{1}{2}} (\theta)$ is the  Wigner d-function.

\subsection{Joint angular distribution for $h\to\Lambda\bar{\Lambda}$ decays} \label{sec:2.1} 

Consider a scalar particle $h$ (such as $\chi_{c0}(1    P)$~\cite{BES:2003kbf}) that couples to $\Lambda\bar{\Lambda}$ through
\begin{align}
    g_{h\Lambda\bar{\Lambda}} h\bar{\Lambda}\Lambda~, \label{lagh}
\end{align}
then, for the two-body decay $h\to\Lambda\bar{\Lambda}$, a straightforward quantum field–theoretic calculation shows that the produced $\Lambda\bar{\Lambda}$ pair is in the helicity state
\begin{align}
   & \left| \Lambda \bar{\Lambda}\right\rangle_{h}=\frac{1}{\sqrt{2}}\left(\left|\frac{1}{2}\right\rangle_\Lambda \left|\frac{1}{2}\right\rangle_{\bar{\Lambda}}-\left|-\frac{1}{2}\right\rangle_\Lambda \left|-\frac{1}{2}\right\rangle_{\bar{\Lambda}}\right)~. \label{lllh}
\end{align}
Accordingly, in the cascade decay $h\to\Lambda(\to p\pi^-)\bar{\Lambda}(\to \bar{p}\pi^+)$, the joint angular distribution of $p$ and $\bar{p}$ is given by
\begin{align}
   \mathcal{W}_h(\theta_1,\theta_2,\phi_1,\phi_2)=\frac{1}{16\pi^2}
   \left(1-\alpha_{\Lambda}\alpha_{\bar{\Lambda}}\cos\theta_1 \cos\theta_2+\alpha_{\Lambda}\alpha_{\bar{\Lambda}}\sin\theta_1 \sin\theta_2\cos\left(\phi_1-\phi_2\right)
   \right)~.\label{hdis}
\end{align}

\subsection{Joint angular distribution for  $a\to\Lambda\bar{\Lambda}$ decays} \label{sec:2.2} 

For a pseudoscalar particle $a$ (such as $\eta_c (1S)$~\cite{BESIII:2025vsr}) that couples to $\Lambda\bar{\Lambda}$ via
\begin{align}
    ig_{a\Lambda\bar{\Lambda}} a\bar{\Lambda}\gamma^5\Lambda~, \label{laga}
\end{align}
the $\Lambda\bar{\Lambda}$ pair produced in the two-body decay $a\to\Lambda\bar{\Lambda}$ is prepared in the helicity configuration
\begin{align}
   & \left| \Lambda \bar{\Lambda}\right\rangle_a=\frac{1}{\sqrt{2}}\left(\left|\frac{1}{2}\right\rangle_\Lambda \left|\frac{1}{2}\right\rangle_{\bar{\Lambda}}+\left|-\frac{1}{2}\right\rangle_\Lambda \left|-\frac{1}{2}\right\rangle_{\bar{\Lambda}}\right)~.\label{llla}
\end{align}
Consequently, for the cascade $a\to\Lambda(\to p\pi^-)\bar{\Lambda}(\to \bar{p}\pi^+)$, the joint angular distribution of $p$ and $\bar{p}$ takes the form
\begin{align}
\mathcal{W}_a(\theta_1,\theta_2,\phi_1,\phi_2)&=\frac{1}{16\pi^2}
   \left(1-\alpha_{\Lambda}\alpha_{\bar{\Lambda}}\cos\theta_1 \cos\theta_2-\alpha_{\Lambda}\alpha_{\bar{\Lambda}}\sin\theta_1 \sin\theta_2\cos\left(\phi_1-\phi_2\right)
   \right)\\
   &=\frac{1}{16\pi^2}
   \left(1-\alpha_{\Lambda}\alpha_{\bar{\Lambda}}\hat{e}_p\cdot\hat{e}_{\bar{p}}
   \right)~.\label{adis}
\end{align}

\section{Local Hidden-Variable Theory Construction} \label{sec:3} 

\subsection{Constraints from angular-momentum conservation and CPT invariance}\label{sec:3.0}

Within an LHVT, angular-momentum conservation implies that, independent of whether the pair originates from a scalar ($h$) or a pseudoscalar ($a$), the spin angular momenta in the c.m. frame satisfy
\begin{align}
    & \vec{S}^\prime_\Lambda+\vec{S}^\prime_{\bar{\Lambda}}=\vec{0}~,
\end{align}
where $\vec{S}^\prime_\Lambda$ and $\vec{S}^\prime_{\bar{\Lambda}}$ denote the spin vectors of $\Lambda$ and $\bar{\Lambda}$, respectively, in the c.m. frame.
Now boost each particle from the c.m. frame to its rest frame. The corresponding boost velocities are $\vec{\beta}_{\Lambda}$ and $\vec{\beta}_{{\bar{\Lambda}}}$, which obey $\vec{\beta}_{\Lambda}+\vec{\beta}_{{\bar{\Lambda}}}=\vec{0}$. Denoting by $\vec{S}_\Lambda$ and $\vec{S}_{\bar{\Lambda}}$ the spin vectors in the respective rest frames, we write
\begin{align}
\vec{S}_{\Lambda/\bar{\Lambda}}=\text{L}\left(\vec{\beta}_{\Lambda/\bar{\Lambda}},\vec{S}^\prime_{\Lambda/\bar{\Lambda}}\right)~.
\end{align}
Because $\vec{S}^\prime_{\bar{\Lambda}}$ and $\vec{\beta}_{\bar{\Lambda}}$ are related to $\vec{S}^\prime_\Lambda$ and $\vec{\beta}_\Lambda$ by a 180-degree rotation about the axis $\vec{S}^\prime_\Lambda\times\vec{\beta}_\Lambda$, spatial rotational invariance implies
\begin{align}
    & \vec{S}_\Lambda+\vec{S}_{\bar{\Lambda}}=\vec{0}~.
\end{align}
Adopting the same reference axes used to parameterize $\hat{e}_p~(\hat{e}_{\bar{p}})$, we describe the direction of $\vec{S}_{\Lambda}$ by
\begin{align}
    & \hat{S}=\hat{S}_\Lambda=-\hat{S}_{\bar{\Lambda}}=\left(\sin x \cos y,\sin x \sin y,\cos x\right)~.\label{Sdirec}
\end{align}
Here, $\hat{S}_\Lambda$ and $\hat{S}_{\bar{\Lambda}}$ denote the unit vectors along $\vec{S}_\Lambda$ and $\vec{S}_{\bar{\Lambda}}$, respectively.

Within the LHVT following~\cite{Bechtle:2025ugc,Abel:2025skj}, let $G(x,y)$ be the joint probability density of $(x,y)$, satisfying
\begin{align}
   & G(x,y)\ge 0~, \quad \int_{-1}^1 d\cos x\int_0^{2\pi}d y~ G(x,y)=1~.\label{normg}
\end{align}
We expand $G(x,y)$ in spherical harmonics,
\begin{align}
    G(x,y)=\sum_{l=0}^{\infty}\sum_{m=-l}^{l}g_{l,m}Y_{l,m}(x,y)~.\label{ggxy}
\end{align}
Since $Y_{0,0}=\frac{1}{\sqrt{4\pi}}$, the normalization condition in Eq.~(\ref{normg}) enforces $g_{0,0}=\frac{1}{\sqrt{4\pi}}$.

In the rest frames of $\Lambda$ and $\bar{\Lambda}$, for fixed $\hat{S}_\Lambda$ and $\hat{S}_{\bar{\Lambda}}$, let $F_\Lambda(\hat{S}_\Lambda\cdot\hat{e}_p)$ and $F_{\bar{\Lambda}}(\hat{S}_{\bar{\Lambda}}\cdot\hat{e}_{\bar{p}})$ denote the single-particle decay angular distributions of $\hat{e}_p$ and $\hat{e}_{\bar{p}}$, respectively, with~\cite{Bechtle:2025ugc,Abel:2025skj}
\begin{align}
   & F_{\Lambda/\bar{\Lambda}}(\hat{S}_{\Lambda/\bar{\Lambda}}\cdot\hat{e}_{p/\bar{p}})\ge 0~,\quad \int_{-1}^1 dz ~F_{\Lambda/\bar{\Lambda}}(z)=\frac{1}{2\pi}~.\label{normf}
\end{align}
Expanding in Legendre polynomials,
\begin{align}
   & F_\Lambda(\hat{S}_\Lambda\cdot\hat{e}_p)=\frac{1}{4\pi}\sum_{l=0}^{\infty} b_l P_l(\hat{S}_\Lambda\cdot\hat{e}_p)~,\label{fl}\\
   & F_{\bar{\Lambda}}(\hat{S}_{\bar{\Lambda}}\cdot\hat{e}_{\bar{p}})=\frac{1}{4\pi}\sum_{l=0}^{\infty} c_l P_l(\hat{S}_{{\bar\Lambda}}\cdot\hat{e}_{\bar{p}})~.\label{flb}
\end{align}
Normalization conditions in Eq.~(\ref{normf}) further impose
\begin{align}
    b_0=c_0=1~.
\end{align}
Using standard spherical-harmonic identities, we have
\begin{align}
   &  P_l(\hat{S}_{\Lambda/\bar{\Lambda}}\cdot\hat{e}_{p/\bar{p}})=\frac{4\pi}{2l+1}\sum_{m=-l}^{l}Y_{l,m}^*(\hat{e}_{p/\bar{p}})Y_{l,m}(\hat{S}_{\Lambda/\bar{\Lambda}})~.\label{fpl}
\end{align}

Employing Eqs.~(\ref{Sdirec}), (\ref{ggxy}), (\ref{fl}), (\ref{flb}), and (\ref{fpl}), we obtain
\begin{align}
   &\mathcal{W}(\theta_1,\theta_2,\phi_1,\phi_2)=\int_{-1}^1d \cos x\int_0^{2\pi}d y~ G(x,y)~F_\Lambda(\hat{S}\cdot\hat{e}_p) ~F_{\bar{\Lambda}}(-\hat{S}\cdot\hat{e}_{\bar{p}})  \label{lhvt1}\\
&=\sum_{l=0}^{\infty}\sum_{k=0}^{\infty}\sum_{m=-l}^{l}\sum_{j=-k}^{k}\frac{b_l c_k (-1)^k}{(2l+1)(2k+1)}~\left\langle Y_{l,m}(\hat{S})~Y_{k,j}(\hat{S})\right\rangle~Y_{l,m}^*(\hat{e}_{p})~Y_{k,j}^*(\hat{e}_{\bar{p}}) \label{lhvt2}\\
&=\sum_{L=0}^{\infty}\sum_{l=0}^{\infty}\sum_{k=0}^{\infty}\sum_{M=-L}^{L}\sum_{m=-l}^{l}\sum_{j=-k}^{k}\frac{g_{L,M}b_l c_k (-1)^k}{(2l+1)(2k+1)}C^{L,l,k}_{M,m,j}~Y_{l,m}^*(\hat{e}_{p})~Y_{k,j}^*(\hat{e}_{\bar{p}})~,\label{lhvt3}
\end{align}
where
\begin{align}
    C^{L,l,k}_{M,m,j}&=\int_{-1}^1d \cos x\int_0^{2\pi}d y~Y_{L,M}(\hat{S})~Y_{l,m}(\hat{S})~Y_{k,j}(\hat{S}) \\
    &=\sqrt{\frac{(2L+1)(2l+1)(2k+1)}{4\pi}}
\begin{pmatrix}
L & l & k \\
0 & 0 & 0
\end{pmatrix}
\begin{pmatrix}
L & l & k \\
M & m & j
\end{pmatrix}~.
\end{align}
$C^{L,l,k}_{M,m,j}$ is nonzero only if all of the following hold:
\begin{itemize}
    \item $|l-k|\le L\le l+k$.
    \item $L+l+k$ is even.
    \item $M+m+j=0$.
\end{itemize}
Using orthogonality of the spherical harmonics (angle brackets denoting expectation values with respect to $\mathcal{W}$), we find
\begin{align}
    & \langle \cos\theta_1 \cos\theta_2\rangle=-\frac{b_1 c_1}{9}  \langle\cos^2 x\rangle~, \label{c1c2}\\
    & \langle \sin\theta_1 \sin\theta_2\cos(\phi_1-\phi_2)\rangle =-\frac{b_1 c_1}{9}  \langle\sin^2 x\rangle~, \label{p1p2} 
\end{align}
and thus,
\begin{align}
    & b_1c_1=-9\langle \hat{e}_p\cdot \hat{e}_{\bar{p}}\rangle~. \label{e1e2}
\end{align}

CPT symmetry requires~\cite{Abel:2025skj}
\begin{align}
    b_l=c_l\times(-1)^l~,\label{bcl}
\end{align}
which implies
\begin{align}
   & F_\Lambda(z)=F_{\bar{\Lambda}}(-z)~.
\end{align}

It should be emphasized that, if angular-momentum conservation is abandoned and $\hat{S}_\Lambda$ and $\hat{S}_{\bar{\Lambda}}$ are assumed to vary independently, namely
\begin{align}
& \hat{S}_{\Lambda/\bar{\Lambda}}=(\sin x_{1/2}\cos y_{1/2},\sin x_{1/2}\sin y_{1/2},\cos x_{1/2})~,  
\end{align}
then, for an arbitrary angular distribution function $\mathcal{W}(\theta_1,\theta_2,\phi_1,\phi_2)$, there always exists the following LHVT realization that is consistent with CPT invariance:
\begin{align}
\mathcal{W}(\theta_1,\theta_2,\phi_1,\phi_2)=&\int_{-1}^1d \cos x_1\int_0^{2\pi}d y_1~\int_{-1}^1d \cos x_2\int_0^{2\pi}d y_2~ G(x_1,x_2,y_1,y_2)\nonumber\\
&\times F_\Lambda(\hat{S}_\Lambda\cdot\hat{e}_p) ~F_{\bar{\Lambda}}(\hat{S}_{\bar{\Lambda}}\cdot\hat{e}_{\bar{p}})~
\end{align}
with
\begin{align}
& F_\Lambda(\hat{S}_\Lambda\cdot\hat{e}_p)=\frac{\delta(\hat{S}_\Lambda\cdot\hat{e}_p-1)}{2\pi}~, \\
& F_{\bar{\Lambda}}(\hat{S}_{\bar{\Lambda}}\cdot\hat{e}_{\bar{p}})=\frac{\delta(\hat{S}_{\bar{\Lambda}}\cdot\hat{e}_{\bar{p}}+1)}{2\pi}~, \\
& G(x_1,x_2,y_1,y_2)=\mathcal{W}(x_1,\pi-x_2,y_1,\pi+y_2)~.
\end{align}

\subsection{LHVT realizability for $h\to\Lambda\bar{\Lambda}$: Impossibility proof}\label{sec:3.1}

Expanding the scalar-induced distribution in spherical harmonics, we obtain
\begin{align}
   &\mathcal{W}_h(\theta_1,\theta_2,\phi_1,\phi_2)=\frac{1}{16\pi^2}
   \left(1-\alpha_{\Lambda}\alpha_{\bar{\Lambda}}\frac{4\pi}{3}\sum_{m=-1}^{1} Y^*_{1,m}(\hat{e}_p)Y^*_{1,-m}(\hat{e}_{\bar{p}})\right)~.
\end{align}
Using Eqs.~(\ref{c1c2})–(\ref{e1e2}), this implies
\begin{align}
    & {b_1 c_1}  \langle\cos^2 x\rangle=\alpha_\Lambda \alpha_{\bar{\Lambda}}~, \label{h1}\\
    & {b_1 c_1}  \langle\sin^2 x\rangle =-2 \alpha_\Lambda \alpha_{\bar{\Lambda}}~. \label{h2}
\end{align}
Since both $\langle\cos^2 x\rangle$ and $\langle\sin^2 x\rangle$ are nonnegative and sum to unity, Eqs.~(\ref{h1}) and (\ref{h2}) cannot be satisfied unless $\alpha_{\Lambda} \alpha_{\bar{\Lambda}}=0$.
According to the latest measurements~\cite{BESIII:2025wxe}, $\alpha_{\Lambda} \alpha_{\bar{\Lambda}}\ne 0$, which implies that $\mathcal{W}_h(\theta_1,\theta_2,\phi_1,\phi_2)$ cannot be reproduced within LHVTs. It is noted that the conclusion here is independent of CPT assumptions.

\subsection{LHVT realizability for $a\to\Lambda\bar{\Lambda}$: Role of CPT symmetry and explicit model}\label{sec:3.2}

Performing a spherical-harmonic expansion, we obtain
\begin{align}
   & \mathcal{W}_a(\theta_1,\theta_2,\phi_1,\phi_2)=\frac{1}{16\pi^2}
   \left(1-\alpha_{\Lambda}\alpha_{\bar{\Lambda}}\frac{4\pi}{3}\sum_{m=-1}^{1}(-1)^m Y^*_{1,m}(\hat{e}_p)Y^*_{1,-m}(\hat{e}_{\bar{p}})\right)~.
\end{align}
Invoking Eqs.~(\ref{c1c2})-(\ref{e1e2}), we find
\begin{align}
    & {b_1 c_1}  \langle\cos^2 x\rangle=\alpha_\Lambda \alpha_{\bar{\Lambda}}~, \\
     & {b_1 c_1}  \langle\sin^2 x\rangle =2 \alpha_\Lambda \alpha_{\bar{\Lambda}}~, \\
     & b_1 c_1 =3\alpha_\Lambda \alpha_{\bar{\Lambda}}~. \label{reque}
\end{align}

When keeping CPT symmetry, can $\mathcal{W}_a(\theta_1,\theta_2,\phi_1,\phi_2)$ be reproduced within an LHVT? The detailed derivations in Appendix~\ref{appen} show that the unique realization with CPT symmetry preserved is
\begin{align}
& F_\Lambda(\hat{S}_\Lambda\cdot\hat{e}_p)=\frac{1+b_1 \hat{S}_\Lambda\cdot\hat{e}_p}{4\pi}~, \\
& F_{\bar{\Lambda}}(\hat{S}_{\bar{\Lambda}}\cdot\hat{e}_{\bar{p}})=\frac{1-b_1 \hat{S}_{\bar{\Lambda}}\cdot\hat{e}_{\bar{p}}}{4\pi}~, \\
& G(x,y)=\frac{1}{4\pi}+\sum_{l=3}^{\infty}\sum_{m=-l}^{l}g_{l,m}Y_{l,m}(x,y)~.\label{gxy}
\end{align}
Note that, because only $b_0$ and $b_1$ are nonzero, one has $C^{L,l,k}_{M,m,j}=0$ for $L\ \ge 3$. Consequently, provided the non-negativity constraint $G(x,y)\ge 0$ is respected, the coefficients $g_{L,M}~(L\ge 3)$ may be chosen arbitrarily without modifying the form of $\mathcal{W}_a(\theta_1,\theta_2,\phi_1,\phi_2)$.
The simplest choice is
\begin{align}
& G(x,y)=\frac{1}{4\pi}~.
\end{align}
Eq.~(\ref{gxy}) also admits discrete measures. For instance, $G(x,y)$ may be nonzero only at the following six points in parameter space, with probability $\frac{1}{6}$ assigned to each:
\begin{align}
   & (\pm 1,0,0)~,\quad (0,\pm 1,0)~,\quad (0,0,\pm 1)~.
\end{align}
However, according to the latest measurements~\cite{BESIII:2025wxe},
\begin{align}
     & b_1^2 =-3\alpha_\Lambda \alpha_{\bar{\Lambda}}>1~.
\end{align}
As a consequence, $F_\Lambda(\hat{S}_\Lambda\cdot\hat{e}_p)$ and $F_{\bar{\Lambda}}(\hat{S}_{\bar{\Lambda}}\cdot\hat{e}_{\bar{p}})$ become negative in part of their domains, demonstrating that, when imposing CPT symmetry, $\mathcal{W}_a(\theta_1,\theta_2,\phi_1,\phi_2)$ cannot be realized within an LHVT.

If CPT symmetry is violated, the coefficients $b_l$ and $c_l$ need not satisfy Eq.~(\ref{bcl}). In that case, can $\mathcal{W}_a(\theta_1,\theta_2,\phi_1,\phi_2)$ be reproduced within an LHVT? We consider the following ansatz:
\begin{align}
& F_\Lambda(\hat{S}_\Lambda\cdot\hat{e}_p)=\frac{1+\sum_{l=1}^{\infty}b_l P_l(\hat{S}_\Lambda\cdot\hat{e}_p)}{4\pi}~, \\
& F_{\bar{\Lambda}}(\hat{S}_{\bar{\Lambda}}\cdot\hat{e}_{\bar{p}})=\frac{1+c_1 \hat{S}_{\bar{\Lambda}}\cdot\hat{e}_{\bar{p}}}{4\pi}~, \\
& G(x,y)=\frac{1}{4\pi}~.
\end{align}
A straightforward calculation shows that reproducing the structure of $\mathcal{W}_a(\theta_1,\theta_2,\phi_1,\phi_2)$ reduces to the single condition $b_1 c_1=3\alpha_{\Lambda} \alpha_{\bar{\Lambda}}$.
\begin{itemize}
    \item Assuming $b_l=0$ for $l>2$,
\begin{align}
 F_\Lambda(z) &=\frac{1+b_1 P_1(z)+b_2 P_2(z)}{4\pi} ~.\label{apb}
\end{align}
According to Appendix~\ref{appenb}, non-negativity of $F_\Lambda(z)~(z\in [-1,1])$ enforces $|b_1|\le \sqrt{3}$. One also has $|c_1|\le 1$. Hence,
\begin{align}
    |\alpha_{\Lambda} \alpha_{\bar{\Lambda}}|=\frac{1}{3}|b_1 c_1|\le \frac{1}{\sqrt{3}}~.
\end{align}
According to the latest measurements~\cite{BESIII:2025wxe},
\begin{align}
    |\alpha_{\Lambda} \alpha_{\bar{\Lambda}}|\approx \frac{1}{\sqrt{3}}~.
\end{align}
Therefore, with $b_{l}=0$ for $l>2$, the empirical value of $|\alpha_{\Lambda} \alpha_{\bar{\Lambda}}|$ sits precisely at the critical bound.
\item We consider an alternative choice for $F_\Lambda(z)$:
\begin{align}
    F_\Lambda(z)=\frac{ue^{u z}}{4\pi \sinh{u}}~.
\end{align}
This $F_\Lambda(z)$ is manifestly non-negative, and its Legendre expansion yields
\begin{align}
    b_0=1~,\quad b_1=-\frac{3}{u}+3\coth{u}\in[-3,3]~.
\end{align}
Therefore, $b_1 c_1~(|c_1|\le1)$ can span the entire interval $[-3,3]$, which covers all admissible values of $\alpha_\Lambda\alpha_{\bar{\Lambda}}$ ($\alpha_\Lambda\alpha_{\bar{\Lambda}}\in[-1,1]$).
\end{itemize}
Therefore, in the presence of CPT violation, $\mathcal{W}_a(\theta_1,\theta_2,\phi_1,\phi_2)$ can indeed be reproduced within an LHVT.

\section{LHVT Realizability for General Spin-Zero $s\to \Lambda\bar{\Lambda}$ Decays}\label{sec:spinzero}

In addition to special spin-zero scenarios with definite parity, it is important to consider the most general spin-zero decay $s\to\Lambda\bar{\Lambda}$. A generic spin-zero state $s$ need not be a pure scalar or pseudoscalar state; its decay amplitude may contain an arbitrary complex mixture of scalar and pseudoscalar couplings. Such a mixture changes the structure of the spin-correlation tensor and may introduce interference terms that are absent in the parity-eigenstate limits. Therefore, the question of LHVT realizability should be addressed directly for the fully general spin-zero decay $s\to\Lambda\bar{\Lambda}$, rather than inferred only from special limiting cases.

In general, the interaction between $s$ and
$\Lambda\bar{\Lambda}$ can be given by
\begin{align}
     s\bar{\Lambda}(g_{1}+ig_{2}\gamma_5)\Lambda~, \label{lagh}
\end{align}
where both $g_1$ and $g_2$ may in general be complex. Then, the polarization state
of the $\Lambda\bar{\Lambda}$ pair produced in the process
$s\to \Lambda\bar{\Lambda}$ can be parametrized as
\begin{align}
   & \left| \Lambda \bar{\Lambda}\right\rangle_s
   =
   \cos z~\left|\frac{1}{2}\right\rangle_\Lambda
   \left|\frac{1}{2}\right\rangle_{\bar{\Lambda}}
   +
   \sin z~e^{i \eta}~
   \left|-\frac{1}{2}\right\rangle_\Lambda
   \left|-\frac{1}{2}\right\rangle_{\bar{\Lambda}}~.
   \label{sstate}
\end{align}
The state $\left| \Lambda \bar{\Lambda}\right\rangle_h$ in Eq.~(\ref{lllh}) and the state $\left| \Lambda \bar{\Lambda}\right\rangle_a$ in Eq.~(\ref{llla}) correspond to $z=-\frac{\pi}{4}$ and $z=\frac{\pi}{4}$, respectively, and both require $\eta=0$.
In the remainder of this Section, we use the free parameters $z$ and $\eta$
rather than $g_1$ and $g_2$ in our discussion. Without loss of generality, we
take $z,\eta\in (-\frac{\pi}{2},\frac{\pi}{2}]$. For completeness of the discussion,
we also treat $\alpha_{\Lambda,\bar{\Lambda}}\in [-1,1]$ as free parameters.

The joint angular distribution of $p$ and $\bar{p}$ is then
\begin{align}
\mathcal{W}_s(\theta_1,\theta_2,\phi_1,\phi_2)
=
\frac{1}{16\pi^2}
\Big(
&1
-\cos(2z)
\left(
\alpha_\Lambda\cos\theta_1
-
\alpha_{\bar\Lambda}\cos\theta_2
\right)
-\alpha_\Lambda\alpha_{\bar\Lambda}
\cos\theta_1\cos\theta_2
\nonumber\\
&-\alpha_\Lambda\alpha_{\bar\Lambda}
\sin(2z)\,
\sin\theta_1\sin\theta_2
\cos(\phi_1-\phi_2+\eta)
\Big)~.
\label{sdis}
\end{align}
Equivalently, the last term may be decomposed into a CP-even part
proportional to $\cos\eta$ and a CP-odd azimuthal structure proportional to
$\sin\eta$. This observation is useful for examining whether the distribution
can be reproduced by an LHVT.

If $\mathcal{W}_s(\theta_1,\theta_2,\phi_1,\phi_2)$ in Eq.~(\ref{sdis}) can
be described within an LHVT, an exact matching to
Eqs.~(\ref{lhvt1}) to (\ref{lhvt3}) necessarily requires
\begin{align}
&\alpha_{\Lambda}\alpha_{\bar{\Lambda}}\sin(2z)\sin{\eta}= 0~.
\label{eta_condition}
\end{align}
This implies that any case with
$\alpha_{\Lambda}\alpha_{\bar{\Lambda}}\sin(2z)\sin{\eta}\ne 0$ cannot be
described by an LHVT, and this conclusion is independent of whether CPT
symmetry is preserved.

In what follows, we impose CPT invariance, namely
$\alpha_{\bar{\Lambda}}=-\alpha_{\Lambda}$ and $b_l=(-1)^l\times c_l$.
The cases $\alpha_{\Lambda,\bar{\Lambda}}=0$ or $\sin(2z)=0$ correspond to
simple situations that can be described within LHVTs and will not be
discussed in detail here. We therefore focus mainly on the nontrivial case
with $\eta=0$ and $\alpha_{\Lambda,\bar{\Lambda}}\ne 0$. Under CPT
invariance, Eq.~(\ref{sdis}) becomes
\begin{align}
\mathcal{W}_s(\theta_1,\theta_2,\phi_1,\phi_2)
=
\frac{1}{16\pi^2}
\Big(
&1
-\alpha_\Lambda \cos(2z)
\left(
\cos\theta_1+\cos\theta_2
\right)
+\alpha_\Lambda^2
\cos\theta_1\cos\theta_2
\nonumber\\
&+\alpha_\Lambda^2\sin(2z)
\sin\theta_1\sin\theta_2
\cos(\phi_1-\phi_2)
\Big)~.
\label{sdis_cpt}
\end{align}
An exact matching between Eq.~(\ref{sdis_cpt}) and
Eqs.~(\ref{lhvt1}) to (\ref{lhvt3}) requires, through an analysis similar to
that in Appendix~\ref{appen}, the following solution:
\begin{align}
& F_\Lambda(\hat{S}_\Lambda\cdot\hat{e}_p)
=
\frac{1+b_1 \hat{S}_\Lambda\cdot\hat{e}_p}{4\pi}~,
\quad
F_{\bar{\Lambda}}(\hat{S}_{\bar{\Lambda}}\cdot\hat{e}_{\bar{p}})
=
\frac{1-b_1 \hat{S}_{\bar{\Lambda}}\cdot\hat{e}_{\bar{p}}}{4\pi}~, \\
& G(x,y)
=
\frac{1}{4\pi}
+
g_{1,0}Y_{1,0}(x,y)
+
g_{2,0}Y_{2,0}(x,y)
+
\sum_{l=3}^{\infty}\sum_{m=-l}^{l}
g_{l,m}Y_{l,m}(x,y)~.
\label{gxy_s}
\end{align}
Here the relevant coefficients are
\begin{align}
& b_1=\pm \alpha_{\Lambda}\sqrt{1+2\sin(2z)}~,
\label{fzb1}\\
& g_{1,0}
=
\mp \frac{1}{2}\sqrt{\frac{3}{\pi}}
\frac{\cos(2z)}{\sqrt{1+2\sin(2z)}}~,
\label{fzg1}\\
& g_{2,0}
=
\frac{1}{2}\sqrt{\frac{5}{\pi}}
\frac{1-\sin(2z)}{1+2\sin(2z)}~.
\label{fzg2}
\end{align}
The signs in Eqs.~(\ref{fzb1}) and (\ref{fzg1}) are correlated. It should be
emphasized that, since $b_{l}=0$ for $l>1$, the values of $g_{l,m}$ with $l>2$ do not affect
the validity of the matching to Eq.~(\ref{sdis_cpt}).

For this solution to be meaningful, one must first require
$1+2\sin(2z)>0$ and $-1\le b_1\le 1$. These conditions lead to
\begin{align}
    z\in \begin{cases}
       (-\frac{\pi}{2},-\frac{5\pi}{12})
       \cup
       (-\frac{\pi}{12},\frac{\pi}{2}]~,
       \quad
       &|\alpha_{\Lambda}| \in (0,\frac{1}{\sqrt{3}}]~,
       \\
       (-\frac{\pi}{2},-\frac{5\pi}{12})
       \cup
       (-\frac{\pi}{12},
       z^\prime]
       \cup
       [\frac{\pi}{2}-z^\prime,
       \frac{\pi}{2}]~,
       \quad
       & |\alpha_{\Lambda}| \in (\frac{1}{\sqrt{3}},1]~.
   \end{cases}
   \label{z_range_b1}
\end{align}
Here, when
$|\alpha_{\Lambda}| \in (\frac{1}{\sqrt{3}},1]$, $z^\prime$ is defined as
\begin{align}
    z^\prime
    =
    \frac{1}{2}
    \sin^{-1}
    \left(
    \frac{1-\alpha_{\Lambda}^2}{2\alpha_{\Lambda}^2}
    \right)~.
\label{zprime}
\end{align}

In addition, the non-negativity of $G(x,y)$ must also be guaranteed, namely
$G(x,y)\ge 0$. If all $g_{l,m}$ with $l>2$ are set to zero, the
non-negativity of $G(x,y)$ requires $z\in
    \left(
    \frac{\sin^{-1}(3-\sqrt{6})}{2},
    \frac{\pi-\sin^{-1}(3-\sqrt{6})}{2}
    \right)$.
However, if appropriate contributions from $g_{l,m}$ with $l>2$ are
taken into account, then the non-negativity condition for $G(x,y)$ can be formulated in terms of the positive semidefiniteness of the matrix
$\mathcal{C}$, given by
\begin{align}
   & \mathcal{C}
   =
   \left(
\begin{array}{cccc}
1 & \mathcal{Q}^{\text{T}}  \\
\mathcal{Q} & \mathcal{J}\\
\end{array}
\right)
=
\left(
\begin{array}{cccc}
1 & 0 & 0 & \frac{\pm \cos (2 z)}{\sqrt{1+2 \sin (2 z)}} \\
0 & \frac{\sin(2 z)}{1+2\sin(2 z)} & 0 & 0 \\
0 & 0 & \frac{\sin(2 z)}{1+2\sin(2 z)} & 0 \\
\frac{\pm \cos (2 z)}{\sqrt{1+2 \sin (2 z)}} & 0 & 0 & \frac{1}{1+2 \sin (2 z)} \\
\end{array}
\right)~,
\label{moment_matrix}
\end{align}
where the first moment $\mathcal{Q}$ and the second moment $\mathcal{J}$ are
defined as
\begin{align}
   & \mathcal{Q}_k=\langle\hat{S}_k\rangle~,
   \quad
   \mathcal{J}_{k,m}=\langle\hat{S}_k\hat{S}_m\rangle~.
\label{moments_def}
\end{align}
Thus, requiring $\mathcal{J}_{i,i}=\langle\hat{S}_i^2\rangle\ge 0$ for $i=1,2,3$ yields
$z\in [0,\frac{\pi}{2}]$.
The same condition can also be obtained directly from the positive semidefiniteness of
$\mathcal{C}$. The eigenvalues $\lambda_i~(i=1,2,3,4)$ of $\mathcal{C}$ are
\begin{align}
    &\lambda_{1,2}
    =
    \frac{\sin(2z)}{1+2\sin(2 z)}~,
    \quad
    \lambda_{3,4}
    =
    \frac{1+\sin (2 z)}{1+2 \sin (2 z)}
    \pm
    \frac{
    \sqrt{
    5 \sin (2 z)+\sin (6 z)-\cos (8 z)+3
    }
    }{
    \sqrt{2} (1+2 \sin (2 z))^{3/2}
    }~.
\label{eigenvalues_C}
\end{align}
Requiring all $\lambda_i~(i=1,2,3,4)$ to be non-negative then gives
\begin{align}
    z\in [0,\frac{\pi}{2}]~.
\label{z_range_C}
\end{align}

\begin{figure}[tbp]
\begin{center}
\includegraphics[width=0.9\textwidth]{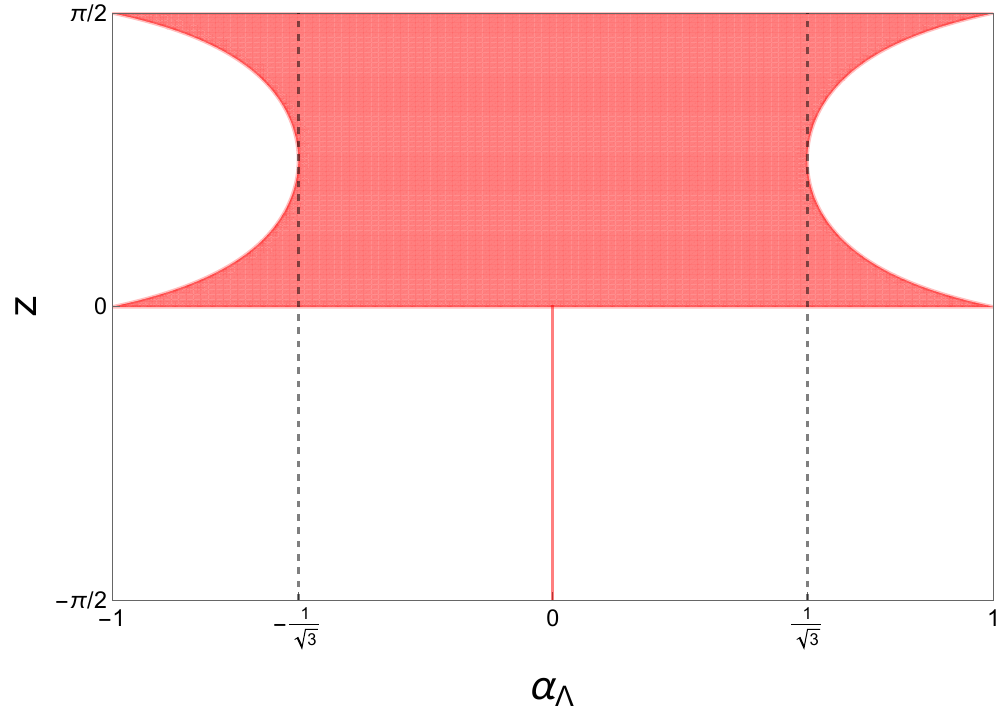}
\end{center}
 \vspace*{-0.1in}
\caption{The red shaded region denotes the parameter region specified by Eq.~(\ref{final_z_range}). 
}\label{dp}
\end{figure}

Combining the constraints from Eqs.~(\ref{z_range_b1}) and
(\ref{z_range_C}), the allowed range of $z$ for which the full solution is
meaningful is
\begin{align}
   z\in \begin{cases}
   (-\frac{\pi}{2},\frac{\pi}{2}]~,
   \quad
   &\alpha_{\Lambda}=0~,
   \\
   [0,\frac{\pi}{2}]~,
   \quad
   &|\alpha_{\Lambda}| \in (0,\frac{1}{\sqrt{3}}]~,
   \\
   [0,z^\prime]\cup [\frac{\pi}{2}-z^\prime,\frac{\pi}{2}]~,
   \quad
   & |\alpha_{\Lambda}| \in (\frac{1}{\sqrt{3}},1]~.
   \end{cases}
\label{final_z_range}
\end{align}
For $\alpha_{\Lambda}=0$, a simple meaningful solution is, for example,
\begin{align}
& F_\Lambda(\hat{S}_\Lambda\cdot\hat{e}_p)
=
F_{\bar{\Lambda}}(\hat{S}_{\bar{\Lambda}}\cdot\hat{e}_{\bar{p}})
=
G(x,y)
=
\frac{1}{4\pi}~.
\label{alpha_zero_solution}
\end{align}
For $\alpha_{\Lambda}\ne 0$, an explicit non-negative realization of
$G(x,y)$ consistent with Eqs.~(\ref{gxy_s}), (\ref{fzg1}), and (\ref{fzg2}) can be chosen as
\begin{align}
G(x,y)
=
\frac{1}{4\pi}
\big(
\delta(\cos x-u_+)
+
\delta(\cos x-u_-)
\big)~.
\label{eq:G-ring}
\end{align}
Here,
\begin{align}
    u_{\pm}
    =
    \frac{-\cos(2z)\pm \sin(2z)}
    {\sqrt{1+2\sin(2z)}}
    ~\text{or}~
    \frac{\cos(2z)\pm \sin(2z)}
    {\sqrt{1+2\sin(2z)}}~.
\label{u_pm}
\end{align}
The sign in front of $\cos(2z)$ is correlated with the sign chosen in
Eqs.~\eqref{fzb1} and \eqref{fzg1}. It can be verified that
$|u_{\pm}|\le 1$ for $z\in [0,\frac{\pi}{2}]$, so that
Eq.~(\ref{eq:G-ring}) indeed defines a non-negative and normalized hidden
spin distribution. Therefore, in the parameter region specified in Eq.~(\ref{final_z_range}) and displayed in Figure \ref{dp}, the QFT angular distribution given in Eq.~(\ref{sdis_cpt}) admits an explicit CPT-invariant LHVT realization.

When $\alpha_{\Lambda,\bar{\Lambda}}= 0$, the joint angular distribution is isotropic and independent of $z$, and an LHVT realization necessarily exists. For $\alpha_{\Lambda,\bar{\Lambda}}\ne 0$, the non-negativity of $G(x,y)$ excludes the possibility of an LHVT realization in the region $z<0$. In the region $z\ge 0$, a smaller value of $|\alpha_{\Lambda,\bar{\Lambda}}|$ makes the joint angular distribution closer to an isotropic one, thereby making an LHVT realization easier to achieve; in particular, all cases with $|\alpha_{\Lambda,\bar{\Lambda}}|\le\frac{1}{\sqrt{3}}$ are LHVT realizable. By contrast, when $|\alpha_{\Lambda,\bar{\Lambda}}|$ is relatively large, namely $|\alpha_{\Lambda,\bar{\Lambda}}|>\frac{1}{\sqrt{3}}$, an LHVT realization becomes easier as the system approaches the unentangled product-state limits, $z=0$ or $\frac{\pi}{2}$. These product states always admit an LHVT realization, independently of the value of $|\alpha_{\Lambda,\bar{\Lambda}}|$.
The latest experimental measurement~\cite{BESIII:2025wxe} gives $\alpha_\Lambda>\frac{1}{\sqrt{3}}$. Therefore, for the $\Lambda\bar{\Lambda}$ pair produced in the general spin-zero decay $s\to \Lambda\bar{\Lambda}$ and parametrized by Eq.~(\ref{sstate}), an LHVT description of the QFT joint angular
distribution is possible, without conflicting with angular-momentum conservation and CPT invariance, only when $\eta=0$ and $z\in [0,z^\prime]\cup [\frac{\pi}{2}-z^\prime,\frac{\pi}{2}]$.

It should be emphasized that, if CPT invariance is not imposed, an exact matching between Eq.~(\ref{sdis}) with $\alpha_{\Lambda,\bar{\Lambda}}\ne 0$ and Eqs.~(\ref{lhvt1}) to (\ref{lhvt3}) requires, in addition to Eqs.~(\ref{fzb1}), (\ref{fzg1}), and (\ref{fzg2}), the following condition:
\begin{align}
   & c_1=\pm \alpha_{\bar{\Lambda}}\sqrt{1+2\sin(2z)}~.
\label{fzb2}
\end{align}
In this case, $b_l$ and $c_l$ with $l>1$, as well as $g_{k,m}$ with $k>2$, are not required to vanish. Rather, they must be chosen such that they do not generate angular structures that are absent in Eq.~(\ref{sdis}).
The requirement $1+2\sin(2z)>0$ excludes the interval $z\in [-\frac{5\pi}{12},-\frac{\pi}{12}]$.
Moreover, the non-negativity of $G(x,y)$ still imposes the constraint $z\in[0,\frac{\pi}{2}]$, which provides a necessary but not sufficient condition for an LHVT realization. A complete analysis of the simultaneous non-negativity conditions for $F_\Lambda(\hat{S}_\Lambda\cdot\hat{e}_p)$, $F_{\bar{\Lambda}}(\hat{S}_{\bar{\Lambda}}\cdot\hat{e}_{\bar{p}})$, and $G(x,y)$ is considerably more involved, due to the possible presence of nonzero $b_l$ and $c_l$ with $l>1$ and $g_{k,m}$ with $k>2$, together with the mutual constraints among these coefficients. 
For this reason, although we have provided a successful LHVT construction for the $a\to \Lambda\bar{\Lambda}$ case ($z=\frac{\pi}{4}$) in a CPT-violating scenario in Sec.~\ref{sec:3.2}, a systematic treatment of the general case without imposing CPT invariance is beyond the scope of the present work and is left for future investigation.

\section{Conclusion} \label{con} 

In this work, we have conducted a comprehensive theoretical analysis of spin entanglement in $\Lambda\bar{\Lambda}$ pairs produced in the two-body decays of spin-zero particles, establishing a clear demarcation between the predictions of QFT and those attainable by LHVTs.

Our investigation yields three principal and sharply contrasting conclusions for different production channels:
\begin{itemize}
    \item For the scalar decay $h \to \Lambda\bar{\Lambda}$, the QFT prediction leads to a unique joint angular distribution, Eq.~(\ref{hdis}), originating from the helicity-entangled state $|\Lambda\bar{\Lambda}\rangle_h$. We have demonstrated that this distribution cannot be reproduced by any LHVT that respects the fundamental constraints of locality and angular-momentum conservation. The derived conditions, $b_1 c_1\langle\cos^2 x\rangle = \alpha_{\Lambda}\alpha_{\bar{\Lambda}}$ and $b_1 c_1\langle\sin^2 x\rangle = -2\alpha_{\Lambda}\alpha_{\bar{\Lambda}}$, are incompatible for non-zero analyzing powers $\alpha_{\Lambda}\alpha_{\bar{\Lambda}} \neq 0$, as confirmed by experimental data. This renders the $h\to\Lambda\bar{\Lambda}$ process a robust and unambiguous probe for ruling out local hidden variables.
    \item For the pseudoscalar decay $a \to \Lambda\bar{\Lambda}$, the situation is more nuanced but equally revealing. The QFT distribution, Eq.~(\ref{adis}), takes the remarkably simple form $1 - \alpha_{\Lambda}\alpha_{\bar{\Lambda}} \hat{e}_p \cdot \hat{e}_{\bar{p}}$. We first show that no CPT-symmetric LHVT can reproduce this result. The requirement $b_1^2 = -3\alpha_{\Lambda}\alpha_{\bar{\Lambda}}$ from the data leads to $|b_1| > 1$, violating the positivity of the single-particle decay functions. Thus, the observation of the QFT distribution in a pseudoscalar decay, combined with the established non-zero analyzing powers, inherently contradicts CPT-symmetric local hidden variable models. Crucially, however, we have constructed an explicit LHVT that successfully reproduces the exact QFT distribution if CPT symmetry is relaxed. This model, characterized by a uniform hidden-variable measure and response functions satisfying $b_1 c_1 = 3\alpha_{\Lambda}\alpha_{\bar{\Lambda}}$, demonstrates that the pseudoscalar channel alone cannot distinguish between QFT and a non-CPT-symmetric LHVT.
    \item
We also consider the most general spin-zero decay
\(s\to\Lambda\bar{\Lambda}\), in which arbitrary complex scalar and pseudoscalar couplings are allowed. The polarization state
of the resulting $\Lambda\bar{\Lambda}$ pair can be written as
$\left| \Lambda \bar{\Lambda}\right\rangle_s
   =
   \cos z~\left|\frac{1}{2}\right\rangle_\Lambda
   \left|\frac{1}{2}\right\rangle_{\bar{\Lambda}}
   +
   \sin z~e^{i \eta}~
   \left|-\frac{1}{2}\right\rangle_\Lambda
   \left|-\frac{1}{2}\right\rangle_{\bar{\Lambda}}$.
Given that the latest experimental result~\cite{BESIII:2025wxe} indicates $\alpha_\Lambda>\frac{1}{\sqrt{3}}$, the QFT joint angular distribution admits an LHVT realization compatible with both angular-momentum conservation and CPT invariance only for $\eta=0$ and $z\in [0,z^\prime]\cup [\frac{\pi}{2}-z^\prime,\frac{\pi}{2}]$, where $z^\prime=
    \frac{1}{2}
    \sin^{-1}
    \left(
    \frac{1-\alpha_{\Lambda}^2}{2\alpha_{\Lambda}^2}
    \right)$.
By contrast, when
$\alpha_{\Lambda}\alpha_{\bar{\Lambda}}\sin(2z)\sin{\eta}\ne 0$ or $z<0$ with $\eta=0$, no LHVT description is possible, irrespective of whether CPT
symmetry is imposed.
\end{itemize}
The scalar channel such as $\chi_{c0}(1P)\to\Lambda\bar{\Lambda}$ serves as a direct test of locality and realism against quantum mechanics, independent of CPT assumptions. In contrast, the pseudoscalar channel such as $\eta_c(1S)\to\Lambda\bar{\Lambda}$ functions as a sensitive probe for CPT symmetry within any local hidden-variable framework; observing the QFT distribution here would force an LHVT to abandon CPT invariance.

These findings furnish clear theoretical benchmarks for ongoing and future experiments at facilities like BESIII, Belle II, and the LHC, which study hyperon-antihyperon pair production. Measuring the joint angular distributions in different production environments will allow for a direct confrontation between QFT and LHVT, potentially delivering the first loophole-free experimental discrimination between these foundational paradigms in the baryon sector.

\appendix
\section{Derivations of the CPT-Symmetric LHVT Model for $a\to\Lambda\bar{\Lambda}$}\label{appen}

With CPT symmetry preserved, we start from the following constraints:
\begin{align}
    & b_0=1~, \\
    & b_1^2 =-3\alpha_\Lambda \alpha_{\bar{\Lambda}}~, \\
    & g_{0,0}=\frac{1}{\sqrt{4\pi}}~.
\end{align}

Any nonzero $g_{1,m}~(m=-1,0,1)$, together with $b_0$ and $b_1$, would induce in $\mathcal{W}_a(\theta_1,\theta_2,\phi_1,\phi_2)$ contributions of the form $Y^*_{1,-m}(\hat{e}_p)Y^*_{0,0}(\hat{e}_{\bar{p}})$ and $Y^*_{0,0}(\hat{e}_p)Y^*_{1,-m}(\hat{e}_{\bar{p}})$, which are absent. Hence $g_{1,m}=0$. The implication chain is
\begin{align}
    g_{1,m},~b_0,~b_1\to Y^*_{1,-m}(\hat{e}_p)Y^*_{0,0}(\hat{e}_{\bar{p}}),~Y^*_{0,0}(\hat{e}_p)Y^*_{1,-m}(\hat{e}_{\bar{p}})\Rightarrow g_{1,m}=0~.
\end{align}
Since $g_{1,0}=0$, terms in $\mathcal{W}_a(\theta_1,\theta_2,\phi_1,\phi_2)$ proportional to $Y^*_{1,m}(\hat{e}_p)Y^*_{1,-m}(\hat{e}_{\bar{p}})~(m=-1,0,1)$ can only originate from $g_{0,0}$ and $g_{2,0}$. Exact coefficient matching then enforces $g_{2,0}=0$. The corresponding chain is
\begin{align}
    & g_{0,0},~g_{2,0},~b_1 \to Y^*_{1,m}(\hat{e}_p)Y^*_{1,-m}(\hat{e}_{\bar{p}}) \Rightarrow g_{2,0}=0~.
\end{align}
Moreover, because $\mathcal{W}_a(\theta_1,\theta_2,\phi_1,\phi_2)$ contains no structures of the form $Y^*_{1,m}(\hat{e}_p)Y^*_{1,j}(\hat{e}_{\bar{p}})~(m+j\ne 0)$, it follows that $g_{2,M}~(M\ne 0)$ must also vanish:
\begin{align}
    & g_{2,M},~b_1 \to Y^*_{1,m}(\hat{e}_p)Y^*_{1,j}(\hat{e}_{\bar{p}}) \Rightarrow g_{2,M}=0~.
\end{align}

Assume now there exists $b_n\ne 0$ for $n>1$. First,
\begin{align}
    & g_{n,m},~b_0,~b_n \to Y^*_{n,-m}(\hat{e}_p)Y^*_{0,0}(\hat{e}_{\bar{p}}),~Y^*_{0,0}(\hat{e}_p)Y^*_{n,-m}(\hat{e}_{\bar{p}}) \Rightarrow g_{n,m}=0~.
\end{align}
In addition, exact coefficient matching implies
\begin{align}
    & g_{n\pm 1,0},~b_1,~b_n \to Y^*_{1,m}(\hat{e}_p)Y^*_{n,-m}(\hat{e}_{\bar{p}}),~Y^*_{n,-m}(\hat{e}_p)Y^*_{1,m}(\hat{e}_{\bar{p}}) \Rightarrow g_{n\pm 1,0}=0~.
\end{align}
Finally, $g_{i,0}$ (with $0\le i\le 2n$ and $i$ even), together with $b_n$, generates terms of the form $Y^*_{n,m}(\hat{e}_p)Y^*_{n,-m}(\hat{e}_{\bar{p}})~(m=0,
\pm 1,\pm 2,...,\pm n)$, and all corresponding coefficients must vanish. Taking into account that $g_{2,0}$ and $g_{n\pm 1,0}$ are forced to zero, a careful count shows there are $n+1$ independent equations, whereas the number of unknowns $g_{i,0}$ is $q(n)$, given by
\begin{align}
    q(n)=
    \begin{cases}
        1~, &\quad n=2,3~,\\
        n-\frac{5-(-1)^n}{2}~, &\quad n>3~.
    \end{cases}
\end{align}
Therefore, $q(n)$ is always smaller than the number of independent constraints, $n+1$.
To preclude the appearance of terms $Y^*_{n,m}(\hat{e}_p)Y^*_{n,-m}(\hat{e}_{\bar{p}})~(m=0,
\pm 1,\pm 2,...,\pm n)$, one must set $b_n=0$ for all $n>1$.

In summary, we conclude that $b_n=0$ for $n>1$ and $g_{l,m}=0$ for $l=1,2$.

\section{Constraint from the Non-Negativity of $F_\Lambda(z)$ with $b_l=0$ for $l>2$}\label{appenb}

Collecting terms in Eq.~(\ref{apb}), we arrive at the following expression:
\begin{align}
F_\Lambda(z) =\frac{b_2}{4\pi}\left(\frac{3}{2}\left(z+\frac{b_1}{3b_2}\right)^2-\frac{3(1-b_2)^2+(b_1^2-3)}{6b_2^2}\right)~.
\end{align}
For the case $b_2>0$, we consider three subcases:
\begin{itemize}
    \item For $-3b_2 \le b_1 \le 3b_2$, the function $4\pi F_\Lambda(z)$ achieves its minimum at $z = -\frac{b_1}{3b_2}$, with the value $-\frac{3(1-b_2)^2+(b_1^2-3)}{6b_2}$. Here, if $|b_1| > \sqrt{3}$, the minimum becomes negative.
    \item For $b_1 > 3b_2$, the minimum of $4\pi F_\Lambda(z)$ occurs at $z = -1$, given by $1 - b_1 + b_2$, and it satisfies the inequality
    \begin{align}
        1-b_1+b_2<1-\frac{2}{3}b_1~.
    \end{align}
    In particular, when $b_1 \ge \frac{3}{2}$, the minimum is negative.
    \item For $b_1 < -3b_2$, the minimum of $4\pi F_\Lambda(z)$ is located at $z = 1$, given by $1 + b_1 + b_2$, and it satisfies
    \begin{align}
        1+b_1+b_2<1+\frac{2}{3}b_1~.
    \end{align}
    In particular, if $b_1 \le -\frac{3}{2}$, the minimum is negative.
\end{itemize}
For $b_2<0$, the function $4\pi F_\Lambda(z)$ attains its minimum either at $z = -1$ with value $1 - b_1 + b_2$ or at $z = 1$ with value $1 + b_1 + b_2$. Under this condition, if $|b_1| \ge 1$, the minimum is negative.

In summary, if $|b_1|> \sqrt{3}$, then $F_\Lambda(z)$ is necessarily negative for some $z\in [-1,1]$. Therefore, to ensure the non-negativity of $F_\Lambda(z)$, one must impose the condition $|b_1|\le \sqrt{3}$.

\begin{acknowledgments}

J. Pei is supported by the National Natural Science Foundation of China (Project No. 12505121), by the Joint Fund of Henan Province Science and Technology R$\&$D Program (Project No. 245200810077), by the Startup Research Fund of Henan Academy of Sciences (Project No. 20251820001),
and by the Scientific and Technological Research Project of Henan Academy of Sciences (Project No. 20262320001).
L Wu is supported in part by the Natural Science Basic Research Program of Shaanxi, Grant No. 2024JC-YBMS-039.
TL is supported in part by the National Key Research and Development Program of China Grant No. 2020YFC2201504, by the
Projects No. 11875062, No. 11947302, No. 12047503,
and No. 12275333 supported by the National Natural Science Foundation of China, by the Key Research
Program of the Chinese Academy of Sciences, Grant
No. XDPB15, by the Scientific Instrument Developing Project of the Chinese Academy of Sciences, Grant
No. YJKYYQ20190049, by the International Partner
ship Program of Chinese Academy of Sciences for Grand
Challenges, Grant No. 112311KYSB20210012, and by
the Henan Province Outstanding Foreign Scientist Studio Project, No.GZS2025008. 

\end{acknowledgments}

\bibliographystyle{jhep}
\bibliography{jhep}

\end{document}